\documentclass[iop]{emulateapj}

\usepackage{natbib}
\usepackage{graphicx}	
\usepackage{amsmath}	
\usepackage{amssymb}	

\newcommand{\beq}{\begin{equation}}
\newcommand{\eeq}{\end{equation}}
\newcommand{\bea}{\begin{eqnarray}}
\newcommand{\eea}{\end{eqnarray}}

\begin{document}

\title{Observational Constraints on Late-Time Radio Rebrightening of GRB-Supernovae}

\author{Charee Peters\altaffilmark{1,2}, 
Alexander J. van der Horst\altaffilmark{3,4},
Laura Chomiuk\altaffilmark{5},
Adithan Kathirgamaraju\altaffilmark{6}
Rodolfo Barniol Duran\altaffilmark{7},
Dimitrios Giannios\altaffilmark{6},
Cormac Reynolds\altaffilmark{8},
Zsolt Paragi\altaffilmark{9},
Eric Wilcots\altaffilmark{1}}

\altaffiltext{1}{Department of Astronomy, University of Wisconsin - Madison, 475 N. Charter Street, Madison, WI 53706-1507, USA}
\altaffiltext{2}{LSSTC Data Science Fellow}
\altaffiltext{3}{Department of Physics, George Washington University, 725 21st Street NW, Washington, DC  20052, USA}
\altaffiltext{4}{Astronomy, Physics, and Statistics Institute of Sciences (APSIS), George Washington University, 725 21st Street NW, Washington, DC  20052, USA}
\altaffiltext{5}{Center for Data Intensive and Time Domain Astronomy, Department of Physics and Astronomy, Michigan State University, 567 Wilson Road, East Lansing, MI 48824, USA}
\altaffiltext{6}{Department of Physics and Astronomy, Purdue University, 525 Northwestern Avenue, West Lafayette, IN 47907, USA}
\altaffiltext{7}{Department of Physics and Astronomy, Sacramento State University, 6000 J Street, M/S 6041, Sacramento, CA 95819}
\altaffiltext{8}{CSIRO Astronomy and Space Science, Kensington, WA 6151, Australia}
\altaffiltext{9}{Joint Institute for VLBI ERIC (JIVE), Oude Hoogeveensedijk 4, NL-7991 PD Dwingeloo, Netherlands}
\email{cpeters@astro.wisc.edu}

\begin{abstract}
We present a search for late-time rebrightening of radio emission from three supernovae (SNe) with associated gamma-ray bursts (GRBs). It has been previously proposed that the unusually energetic SNe associated with GRBs should enter the Sedov-Taylor phase decades after the stellar explosion, and this SN ``remnant'' emission will outshine the GRB radio afterglow and be detectable at significant distances. We place deep limits on the radio luminosity of GRB\,$980425$/SN\,$1998$bw, GRB\,$030329$/SN\,$2003$dh and GRB\,$060218$/SN\,$2006$aj, $10-18$ years after explosion, with our deepest limit being $L_{\nu}$ $< 4 \times 10^{26}$ erg s$^{-1}$ Hz$^{-1}$ for GRB\,$980425$/SN\,$1998$bw. We put constraints on the density of the surrounding medium for various assumed values of the microphysical parameters related to the magnetic field and synchrotron-emitting electrons. For GRB\,$060218$/SN\,$2006$aj and GRB\,$980425$/SN\,$1998$bw, these density limits have implications for the density profile of the surrounding medium, while the non-detection of GRB\,$030329$/SN\,$2003$dh implies that its afterglow will not be detectable anymore at GHz frequencies. 
\end{abstract}



\section{Introduction} \label{s:1}
Due to the extreme luminosities of gamma-ray bursts (GRBs), releasing $\sim 10^{51}$ erg  of kinetic energy on a timescale of seconds, they can be detected out to very large redshifts and provide a unique way to study physics in extreme conditions \citep[e.g.,][]{vanParadijs2000,Frail2001,Meszaros2002,Gehrels2009}. GRBs can emit long-lasting synchrotron emission spanning X-ray to radio frequencies, known as afterglows \citep{Costa1997,VanParadijs1997,Frail1997}. The radio emission of a GRB, sometimes spanning hours to years after the initial outburst, is dominated by this synchrotron emission produced by the GRB ejecta, which start out as a collimated jet and gradually spread to expand more isotropically \citep[e.g.,][]{Rhoads1999,Frail2000}. When the ejecta interact with the ambient medium, they amplify the magnetic field and accelerate particles to relativistic speeds \citep{Sari1998}. Radio observations play an essential role in understanding GRB afterglows, as they provide information about the energetics of the explosion, the ambient medium, shock physics, and the relativistic expansion velocity in the jets (for a review, see \citealt{Granot2014}). 

Jets are not the only ejecta expelled in the GRB event; if there is an associated supernova (SN), a spherical outflow is also present. Long-duration GRBs, unlike short GRBs, have been found to have associated SNe and may provide insights into the deaths of massive stars \citep{Galama1998,Hjorth2003}. The SNe associated with GRBs are of Type~Ic and feature broad lines in their optical spectra, implying fast moving ejecta with velocities $\sim0.1$ times the speed of light~$c$ \citep{Mazzali2007,Hjorth2012}. These broad-line Type~Ic SNe display kinetic energies that are $\sim$10 times greater than those of GRBs or normal Type~Ic SNe ($\sim10^{52}$ erg; \citealt{Matheson2001, Woosley2006, Drout2011, Hjorth2012, Melandri2014}). 

Although the SN ejecta have a slower maximum velocity, they are much more massive ($\sim 1 - 12$ M$_{\odot}$, \citealt{Zhang2018, Taubenberger2011, Tomasella2018}) than GRB jets ($\sim 10^{-6}$ M$_{\odot}$; \citealt{Panaitescu2002}) and are expected to coast a longer time before decelerating. After the explosion, the SN ejecta will remain in free expansion for a few decades, and will sweep up material from the surrounding medium. The SN ejecta interact with the surrounding medium, accelerating particles to relativistic speeds and amplifying the magnetic field, producing radio synchrotron emission much like in a typical SN remnant (\citealt{Dubner2015}; \citealt{Duran2015a, Kathirgamaraju2016}). 

This radio emission peaks when the SN has swept up an equivalent mass to the initial ejected mass, at the Sedov-Taylor time \citep{Taylor1950, Sedov1959}. For typical SNe, the Sedov-Taylor time is $\sim1,000$ years after the explosion (e.g., \citealt{Berezhko2004}). The Sedov-Taylor time may be $\sim$2 orders of magnitude shorter for the more energetic GRB/SNe, due to their large expansion velocities \mbox{\citep{Duran2015a}}. Because of the likeness to typical SN remnants (SNRs), we refer to GRB/SN radio emission on decades-long scales as ``SNR emission" throughout the rest of this paper. After peaking at the Sedov-Taylor time, the radio emission will decline throughout the Sedov-Taylor phase, as the SNR blast wave decelerates \citep{Berezhko2004, Duran2015a, Kathirgamaraju2016, Sarbadhicary2017}.  

In the first few years after a GRB/SN explosion, while the SN ejecta coast in a free expansion phase, the GRB shock decelerates from ultra-relativistic to non-relativistic speeds. Around $10$ years after the burst, the radio emission from the SN shock approaches the same prominence as the emission from the GRB shock. Ultimately, the SNR emission dominates the total emission due to its higher kinetic energy. \cite{Duran2015a} and \cite{Kathirgamaraju2016} modeled the GRB afterglow and SNR emission, and find that SNRs accompanying nearby GRBs should become detectable with sensitive modern radio telescopes some $\sim$20--50 years after explosion
(assuming that the GRB/SN is interacting with a $1$ cm$^{-3}$ medium, and is at a nearby distance of z $\lesssim 0.2$; see \citealt{Duran2015a}). 
Here we present our search for the radio SNRs associated with three nearby, well-studied GRB/SNe: GRB\,980425/SN\,1998bw, GRB\,$030329$/SN\,$2003$dh, and GRB\,060218/SN\,2006aj. Radio emission on timescales from days to years has previously been detected in these GRB/SN systems. 

Detecting the radio rebrightening of a GRB/SN would mark the first time that we have watched a SN transform into an SNR.  Although there have been efforts to detect the radio re-brightening which defines the start of the SNR phase decades after SN explosion (e.g., \citealt{Stockdale2006,Dittmann2014}), this has yet to be done successfully. Our study focuses on sources that should reach the Sedov-Taylor time faster and have higher luminosity, compared to SNe with more typical energetics (i.e. $\sim 10^{51}$ erg). Indeed, GRB/SNe may present some of the best prospects for detecting this rebrightening radio emission, despite being much further away than the SNe in the \citet{Stockdale2006} sample. Detection of the SNR radio emission would also develop our understanding of particle acceleration and magnetic field amplification in $\sim0.1\,c$ shocks. Finally, the radio rebrightening offers a chance to study the properties of a GRB/SN system, including constraining the density of the ambient medium and potentially settling if the SNe associated with GRBs truly are extra-energetic. 

In Section \ref{s:2}, we describe the models for radio SNR emission from a GRB/SN. Section \ref{s:3} introduces our sample of three GRB/SN sources, and the observations of these sources are described in Section \ref{s:4}. The results of the observations and analysis are discussed in Section \ref{s:5}, and our conclusions are laid out in Section \ref{s:6}.

\section{Models of early SNR radio emission} \label{s:2}

When we began this study of GRB/SN systems, we followed the predictions by \citet{Duran2015a} to estimate radio emission from the SN ejecta at the time of our observations ($\sim10-20$ years after explosion). While our observations were being conducted, more sophisticated models of the SNR emission from GRB/SN systems were published \citep{Kathirgamaraju2016}. Within this paper, we use our deep limits and the models of \citet{Kathirgamaraju2016}, which build on the models of \citet{Duran2015a,Duran2015b}, to constrain parameters which determine the SNR radio luminosity. Here we outline the details of the calculations and models of \citet{Kathirgamaraju2016}. We discuss the implications of these models with our observational limits in more detail in Section \ref{s:5}.

The flux from an SNR reaches a maximum flux density ($F_p$) at the deceleration time of the SN ($t_{\rm dec}$; essentially, the Sedov-Taylor time). The equations for these quantities are given in \citet{Duran2015a} as:
\beq
t_{\rm dec} \approx 29\, \beta_{\rm SN,-1}^{-5/3}\, (E_{\rm SN,52.5}/n_0)^{1/3}\,(1+z) \ \rm yr
\eeq
\beq \label{eq:2}
\begin{split}
F_p \approx & \, 440\, \bar{\epsilon}_{e,-1}\, \epsilon_{\text{B},-2}^{\frac{1+p}{4}}\ \, \beta_{\rm SN,-1}^{\frac{1+p}{2}}\, 
E_{\rm SN,52.5}\, n_0^{\frac{1+p}{4}}\, \nu_{\rm GHz}^{\frac{1-p}{2}} \\ & 
\times (1+z)^{\frac{1-p}{2}}\, d_{\rm 27}^{-2}\ \mu {\rm Jy}
\end{split}
\eeq
Here, $\epsilon_e$ and $\epsilon_B$ are the fraction of the post-shock energy transferred to the relativistic electrons and amplified magnetic field, respectively (in the above equation, they are scaled to convenient values, $\epsilon_{\text{B},-2} = \epsilon_{\text{B}}/0.01$ and $\epsilon_{\text{e},-1} = \epsilon_{\text{e}}/0.1$). The power-law index of the non-thermal electron energy distribution accelerated by the SN blast wave is $p$, and  $\bar{\epsilon}_{e,-1} \equiv 4 \epsilon_{\text{e},-1} (p-2)/(p-1)$. The volume number density of the external medium is $n_0$ in units of cm$^{-3}$. $E_{\rm SN, 52.5}$ is the energy of the SN normalized to $10^{52.5}$ erg, and $\beta_{\rm SN, -1}$ is the ratio of the mass-averaged speed of the SN ejecta and the speed of light ($v/c$) normalized to 0.1. The observing frequency in GHz is $\nu_{\rm GHz}$, $d_{\rm 27}$ is the luminosity distance normalized to $10^{27}$ cm, and $z$ is the redshift. The above equations assume we are observing within max($\nu_a,\nu_m$) $< \nu < \nu_c$, with $\nu_a$, $\nu_m$, and $\nu_c$ being the self-absorption frequency, peak frequency, and cooling frequency, respectively. We note that the numerical pre-factor of equation \ref{eq:2} has a $p$ dependence and $p=2.5$ has been assumed. We do not include the pre-factor's $p$-dependence in Equations 2, 4, or 5 for simplicity. The effect of changing $p$ on this pre-factor can be easily incorporated, and we refer interested readers to the work of \cite{Sironi2013} for more information. Throughout the analysis in this paper, We assume $p=2.5$, but leave $p$ in our equations for readers to see where this dependence occurs.

We assume the ejecta are homologously expanding with a range of velocities, meaning that the velocity of the ejecta is linearly proportional to the radius (i.e., the fastest moving ejecta are outermost, while the inner ejecta expand slowest). The ejecta expanding with $\beta = v/c$ have associated energy $E$, this energy is distributed as $E\propto(\beta\gamma)^{-\alpha}$ for $\beta\ge \beta_{\rm SN}$, and the integrated energy distribution is normalized to $E_{\rm SN}$. Here, we take $\alpha = 5$, consistent with the theory for outer SN ejecta \citep{Matzner99, Tan01}.

\begin{figure*}[!t]
\centering
 \includegraphics[trim=0 0 0 0, clip, width=7in]{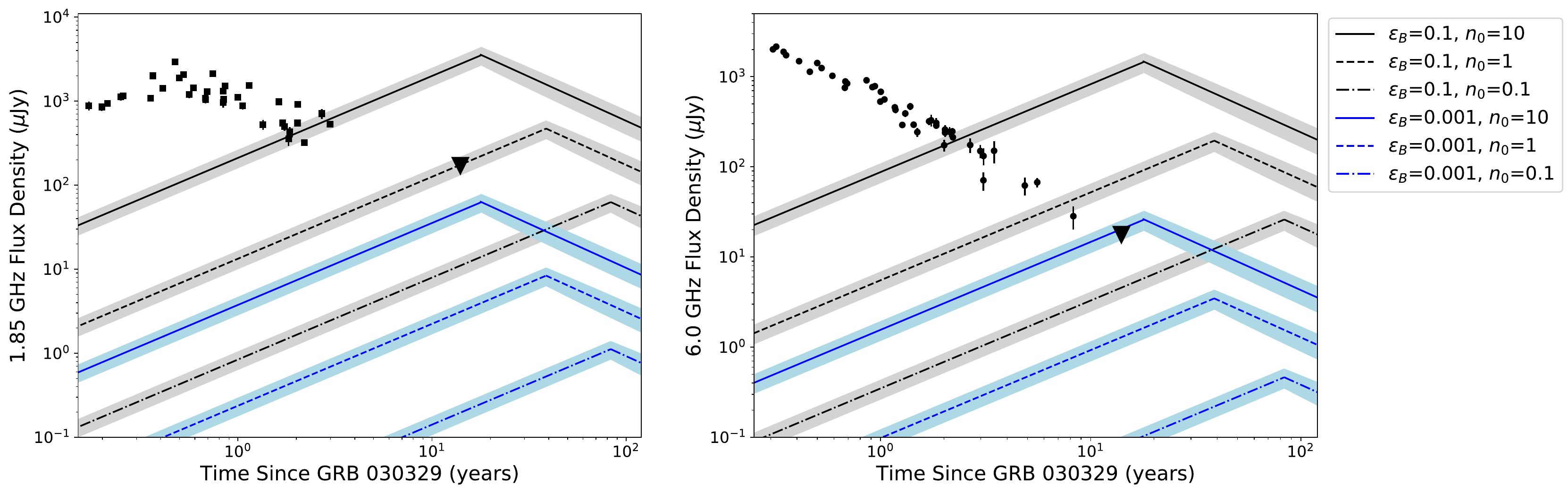}
 \caption{ Radio light curves of  GRB\,$030329$/SN\,$2003$dh at $1.85$ GHz (left) and $6$ GHz (right), including the published GRB afterglow and our late-time upper limits (black triangles). Model radio light curves of the radio SNR rebrightening are superimposed as blue and grey bands; each panel shows six models with varying $\epsilon_B$ and $n_0$. For $\epsilon_B$, we adopt values of $0.1$ (grey lines) and $10^{-3}$ (blue lines). For the ambient density, we adopt values of $n_0 = 0.1$ cm$^{-3}$ (dot-dashed lines), $n_0 = 1$ cm$^{-3}$ (dashed lines), and $n_0 = 10$ cm$^{-3}$ (solid lines). For these models, we use the SN energy and velocity of SN\,$2003$dh given in Table 1 (the bands surrounding each model line represent the uncertainties in the SN energy), and assume $\epsilon_e = 0.1$, $p = 2.5$ and $\alpha = 5$. GRB\,$030329$ afterglow data (filled circles and squares) are from \citet{Berger2003, Frail2005,Resmi2005,vanderHorst2005,vanderHorst2008, Mesler2012}.
\label{Fig_GRB03LightCurveModels}}
\end{figure*}

The radio light curve of the SN both before and after $t_{\rm dec}$ can be expressed as $F_{\nu} = F_p (t/t_{\rm dec})^{-s}$, where $s$ depends on the energy distribution of mass in the SN blast wave. Then $s$ becomes 
\begin{equation}\label{s2}
    s=
    \begin{cases}
      \frac{15p-21-6\alpha}{10+2\alpha}, & t < t_{\rm dec}; \text{non-relativistic phase} \\
      \frac{3(1+p)-6\alpha}{10+2\alpha}, & t < t_{\rm dec}; \text{deep Newtonian phase}\\
      \frac{3(1+p)}{10}, & t > t_{\rm dec}
    \end{cases}
 \end{equation}
These expressions for $s$ and the corresponding fluxes are given in \citet{Kathirgamaraju2016}. The deep Newtonian phase sets in when the speed of the blast wave decreases to $\beta_{\rm DN} = 0.2 {{\bar\epsilon_{e,-1}}}^{-1/2}$, a few years after explosion \citep{Sironi2013}. 

During the deep Newtonian phase, when $t_{\rm DN}<t <  t_{\rm dec}$, the flux density in $\mu$Jy increases with time as:
\beq \label{eq:4}
\begin{split}
F_{\nu} \approx  & \, 440 \, \bar{\epsilon}_{e,-1} \, \epsilon_{\text{B},-2}^{\frac{1+p}{4}} \, \beta_{\rm SN,-1}^{\frac{\alpha(11+p)}{10+2\alpha}} \, E_{\rm SN,52.5}^{\frac{11+p}{10+2\alpha}}  \, n_0^{\frac{3+5\alpha+p(3+\alpha)}{20+4\alpha}}\\
& \times\nu_{{\rm GHz}}^{\frac{1-p}{2}} \, (1+z)^{\frac{8-5\alpha-p(\alpha+2)}{10+2\alpha}} \, d_{\rm 27}^{-2} \, \Big(\frac{t}{29 \, {\rm yr}}\Big)^{\frac{6\alpha-3(1+p)}{10+2\alpha}}
 \end{split}
 \eeq 
 If $t$ is greater than $t_{\rm dec,SN}$, the flux density decreases with time as:
 \beq \label{eq:5}
 \begin{split}
F_{\nu} \approx & \, 440 \, \bar{\epsilon}_{e,-1} \, \epsilon_{\text{B},-2}^{\frac{1+p}{4}} \ \, E_{\rm SN,52.5}^{\frac{11+p}{10}}  \, n_0^{\frac{3(1+p)}{20}}\nu_{{\rm GHz}}^{\frac{1-p}{2}}(1+z)^{\frac{4-p}{5}} \\ & \times d_{\rm 27}^{-2} \, \Big(\frac{t}{29 \, {\rm yr}}\Big)^{\frac{-3(1+p)}{10}}
\end{split}
 \eeq 
 
In this paper, we create model light curves by melding together  equations \ref{eq:4} and \ref{eq:5} at their intersection point, which is approximately the Sedov-Taylor time. The combination of these equations is what we refer to as our model radio light curves and are shown in Figures 1, 2, and 3. Each figure assumes the appropriate explosion parameters for the observed GRB/SN (Table 1). The range of model parameter values ($\epsilon_B$ and $n_0$) are chosen specifically to allow the figures to center and focus on the observational upper-limits for each GRB/SN. We explore a larger parameter space in Figure \ref{Fig_GRB98SetFluxPlots}. Given our observation times in this paper, we typically expect to be in the deep Newtonian phase ($t_{\rm DN}<t<t_{\rm dec}$), hence equation \ref{eq:4} would be the relevant equation for the flux density evolution.
  
We can use this theoretical framework to constrain parameters like $\epsilon_e$, $\epsilon_B$, $E_{\rm SN}$, and $n_0$ from a measurement of $F_{\nu}$. Rearranging equation \ref{eq:4} and taking $\alpha=5$, we find
 \beq \label{eps_vs_n}
 \begin{split}
 \bar{\epsilon}_{e,-1} \, \epsilon_{\text{B},-2}^{\frac{1+p}{4}}\, n_0^{\frac{7+2p}{10}} \lesssim  &\left(\frac{F_\nu}{440\, \mu {\rm Jy}}\right)\left(\frac{t}{29\, {\rm yr}}\right)^{\frac{3(p-9)}{20}}\beta_{\rm SN,-1}^{-\frac{(55+5p)}{20}}\\
 & \times E_{\rm SN,52.5}^{-\frac{(11+p)}{20}}\, \nu_{{\rm GHz}}^{\frac{p-1}{2}}\, (1+z)^{\frac{17+7p}{20}}\, d_{\rm 27}^{2}
 \end{split}
 \eeq
Figure \ref{Fig_GRB98SetFluxPlots} demonstrates the degeneracies between $\epsilon_B$, $E_{SN}$, and $n_0$, given observational constraints on GRB\,$980425$/SN\,$1998$bw. We discuss Figures 1--4 in more detail in \S 5.

\section{Sample} \label{s:3}
In an effort to detect the predicted radio emission from the GRB/SN remnant, we considered the sample of long GRBs with associated SNe. We calculated which of these were most likely to show a detectable radio re-brightening using the methods outlined in \S 3.1 of \cite{Duran2015a}. We took sources that had a time since explosion of over 10 years and were nearby ($z < 0.1$). This narrowed the pool down to two events: GRB\,$980425$/SN\,$1998$bw and GRB\,$060218$/SN\,$2006$aj. We also included GRB\,$030329$/SN\,$2003$dh, which is also $>$10 years old but further away ($z = 0.17$), as it is a very well-monitored event with the longest radio afterglow ever detected \citep{vanderHorst2008,Mesler2013}. For all other GRB/SNe, the radio SNR is predicted to be faint due to the explosion being either too distant or too recent \cite{Duran2015a}. Our target sample can be found in Table 1, along with redshifts, SN ejecta velocities at $\sim10$ days after explosion \citep{Mazzali2007}, and SN energies. 

\begin{deluxetable}{lcccc}[!b]
\tablecaption{Basic Data on Target GRB/SNe \label{SampleTable}}
\tablehead{GRB & SN & $z$ & $\upsilon$ [$10^3$ km/s] & $E_{SN}$ [$10^{51}$ erg]}
\startdata
   $030329$ & 2003dh & $0.1685^{a}$ & $29 \pm 5.8^{d}$ & $40 \pm 10^{e}$ \\
   $060218$ & 2006aj & $0.0335^{b}$ & $19 \pm 3.8^{d}$ & $2 \pm 0.5^{f}$ \\
   $980425$ & 1998bw & $0.0083^{c}$ & $24 \pm 4.8^{d}$ & $50 \pm 5^{g*}$

\enddata
\tablenotetext{a}{\citet{Greiner2003}; $^{b}$\citet{Mirabal2006}; $^c$\citet{Lidman1998}; $^d$\citet{Mazzali2007b} errors are taken as $20\%$; $^{e}$\citet{Mazzali2003}; $^f$\citet{Mazzali2006}; $^g$\citet{Iwamoto1998, Mazzali2007}; $^*$Value confirmed via private communication with Mazzali.}
\end{deluxetable}

\begin{deluxetable*}{cccccccc}[!b]
\tablecaption{Log of Radio Observations\label{ObsTable}}
\tablehead{Source & RA & Dec & Telescope & UT Date & Band \& Central & Bandwidth & Time on \\
   & (h:m:s) & ($^{\circ}:^{\prime}:^{\prime\prime}$) & & Observed & Frequency & (GHz) & Source}
\startdata
GRB\,$030329$/ & 10:44:50.02 & $+$21:31:18.10 & VLA & 2016 Mar $23$ & C-band ($6$ GHz) & $4$ GHz & $30$ min \\
SN\,$2003$dh & & & (C-config) & & L-band ($1.5$ GHz) & $1$ GHz & $54$ min \\
 & & & & & & & \\
GRB\,$060218$/ & 03:21:39.67 & $+$16:52:02.20 & VLA & 2016 Feb $18$ & C-band ($6$ GHz) & $4$ GHz & $30$ min \\
SN\,$2006$aj & & & (C-config) & & L-band ($1.5$ GHz) & $1$ GHz & $54$ min\\
 & & & & & & & \\
GRB\,$980425$/ & 19:35:03.17 & $-$52:50:46.1 & LBA & 2015 Nov $16$ & L-band ($1.65$ GHz) & $32$ MHz & $320$ min\\
SN\,$1998$bw & & & & & & & \\
\enddata
\end{deluxetable*}

\begin{figure*}[!t]
 \includegraphics[trim=0 0 0 0, clip, width=7in]{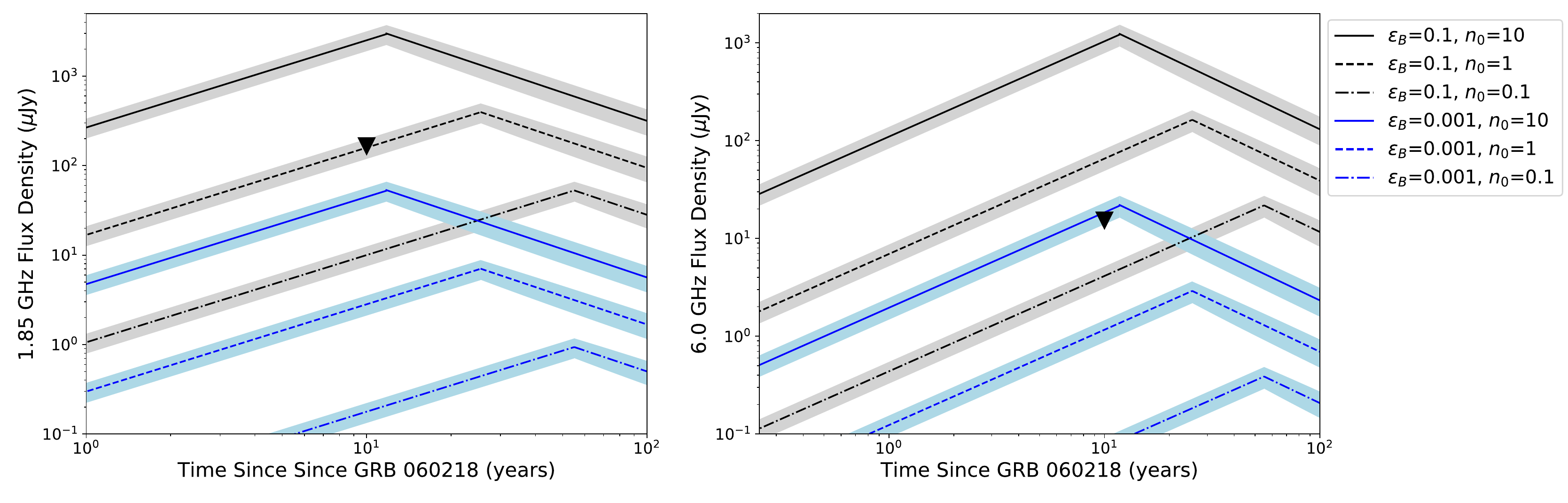}
 \caption{Radio upper limits (black triangles) for GRB\,060218/SN\,2006aj at $1.85$ GHz (left) and $6$ GHz (right), overplotted on model radio light curves.  The light curve models, assumptions, and symbols are the same as for Figure 1, except we use the SN energy and velocity of SN\,2006aj from Table 1. Observations of the GRB\,060218 afterglow are not shown, as the GRB flux decreased rapidly and would not be visible over the timescales featured in this plot.}
\label{Fig_GRB06LightCurveModels}
\end{figure*}

\textbf{GRB\,030329/SN\,2003dh} is at a low redshift, $z = 0.1685$ \citep{Greiner2003}, for a GRB with typical luminosity (compared to the low-luminosity GRBs that are often detected at such low redshifts). It is one of the most well-studied GRB afterglows with radio coverage between $0.64$ and $95$ GHz from only a half day out to almost a decade after the initial gamma-ray detection \citep{Berger2003, Frail2005,  Resmi2005, vanderHorst2005, vanderHorst2008, Mesler2012}. We show the radio observations of the afterglow in Figure \ref{Fig_GRB03LightCurveModels}. Original data were obtained at $1.4$ and $4.9$~GHz; we scaled them respectively to our observing frequencies of $1.85$ and $6$~GHz using the observed radio spectral index, $F_{\nu} \propto \nu^{-0.54}$ \citep{vanderHorst2008}. 
VLBI observations complement the radio light curve data and provide measurements of the source size and evolution \citep{Taylor2004, Taylor2005, Pihlstr2007}. There has also been detailed optical study of the associated SN\,2003dh \citep[e.g.,][]{Stanek2003}, which had a high kinetic energy of $4 \times 10^{52}$ erg \citep{Mazzali2003}.

\textbf{GRB\,060218/SN\,2006aj} is located at a redshift of $z = 0.0335$ \citep{Mirabal2006}. Despite being closer than GRB\,030329, it had a lower intrinsic luminosity \citep{Cobb2006}. The GRB radio afterglow has been detected, but was faint and had sparse sampling over time \citep{Soderberg2006, Kaneko2007}. Therefore the GRB blast wave parameters are not well constrained. However, SN\,$2006$aj was well-studied at optical wavelengths (e.g., \citealt{Sollerman2006}), yielding a measurement of the SN kinetic energy substantially lower than SN\,$2003$dh ($2 \times 10^{51}$ erg; \citealt{Mazzali2006}).  

\textbf{GRB\,980425/SN\,1998bw}
The third GRB with associated SN in our sample is GRB\,$980425$/SN\,$1998$bw, at a redshift of $0.0083$ \citep{Lidman1998}. This was the first GRB found to have an associated SN \citep{Galama1998}, and a radio counterpart was well-detected and monitored at radio wavelengths \citep{Kulkarni1998}, although we do not show it in Figure 3 because it had faded by 1 year after explosion \citep{Frail2003}. Modeling of the GRB data is consistent with a viewing angle misaligned with the GRB jet axis \citep{Ioka2001, Yamazaki2003}. Optical observations and modeling of SN\,$1998$bw imply a kinetic energy comparable to SN\,$2003$dh, $5 \times 10^{52}$ erg \citep{Iwamoto1998}.

\section{Observations} \label{s:4}
In order to detect the radio re-brightening of GRB/SNe, we observed the three objects described in Section \ref{s:3} in 2016. Both GRB\,$030329$/SN\,$2003$dh and GRB\,$060218$/SN\,$2006$aj were observed using the NSF's Karl G.\ Jansky Very Large Array (VLA). GRB\,$980425$/SN\,$1998$bw is a southern source not visible to the VLA, and the host galaxy has bright radio emission, so we observed it with the Australian Long Baseline Array (LBA; Table 2). 

Below is a summary of the observations conducted at each telescope. Each GRB/SN was observed in L band ($1$ -- $2$ GHz) and both GRB\,$030329$/SN\,$2003$dh and GRB\,$060218$/SN\,$2006$aj were observed in C band ($4$ -- $8$ GHz). These particular bands were chosen as a trade-off between sensitivity, resolution, and brightness. Higher frequencies (e.g., C band) are more sensitive than lower frequencies at the VLA, and provide higher resolution on the GRB/SN while resolving out the host galaxy flux. However, the GRB/SNe should be emitting optically-thin synchrotron, and should therefore be brighter at lower frequencies (e.g., L band). To improve our chances of detection, we therefore observe in both L and C bands with the VLA, as the image resolution is limited. Observations with the LBA are much higher resolution but are limited in sensitivity, so in this case we focus on L band observations.

\begin{deluxetable*}{lccccc}
\tablecaption{Summary of Measurements from Radio Observations   \label{MeasurementsTable}}
\tablehead{Source & Frequency & Image r.m.s. & 3$\sigma$ Upper Limit & Time since SN & Luminosity$^{a}$ \\
    & (GHz) & ($\mu$Jy/beam) &($\mu$Jy) & (yr) & ($10^{27}$ erg s$^{-1}$ Hz$^{-1}$)}
 \startdata
GRB\,$030329$/ & $6$ & $5.0$ & $17.6$ & 13.0 & $<13.8$ \\
SN\,$2003$dh & $1.85$ & $55$ & $170$ & 13.0 & $<133$ \\
    & $1.22$ & $80$ & $245$ & 13.0 & $<192$ \\
    & & & & & \\
GRB\,$060218$/ & $6$  & $5.1$ & $15.3$ & 10.0 & $<0.4$ \\
SN\,$2006$aj & $1.85$ & $55$ & $165$ & 10.0 & $<4.1$ \\
     & $1.22$ & $91$ & $281$ & 10.0 & $<6.9$ \\
    & & & & & \\
GRB\,$980425$/ & $1.65$  & $90$ & $270$ & 17.6 & $<0.42$ \\
    SN\,$1998$bw & & & & & 
\enddata  
\tablenotetext{a}{3$\sigma$ limit on the spectral luminosity assuming redshifts listed in Table 1.}
\end{deluxetable*}

\subsection{VLA Observations}
During the VLA's $2016$ C-configuration, we observed GRB\,$030329$/SN\,$2003$dh and GRB\,$060218$/SN\,$2006$aj with the VLA (Program ID VLA/16A-309). Both GRB/SNe were observed for $54$ minutes in L band ($1$ -- $2$ GHz) and $30$ minutes in C band ($4$ -- $8$ GHz). The L-band observations had $16$ spectral windows with a width of $64$~MHz each.  The C-band observations had $32$ spectral windows, each $128$~MHz wide. All spectral windows were sampled with 64 channels, and all observations were carried out in full polarization mode.

For GRB\,$030329$/SN\,$2003$dh, we used 3C286 as the flux calibrator and J1103+2203 as the phase calibrator. For GRB\,$060218$/SN\,$2006$aj, we used 3C147 as the flux calibrator and J0318+1628 as the phase calibrator. The data were edited and reduced using standard routines in both AIPS and CASA \citep{McMullin2007,Greisen2003}. Images were created in AIPS, using a Briggs Robust value of 0. We split the data from each receiver band into two or more frequency chunks and imaged them separately, to assuage imaging artifacts borne of the large fractional bandwidths. As all images yield non-detections, in each receiver band we smoothed the higher-frequency image to the resolution of the lower-frequency image, and then averaged the images together using appropriate noise-based weights in AIPS' \verb|comb|.

GRB\,$060218$/SN\,$2006$aj has a very bright source less than a degree away ($\sim$8~Jy at 1.4~GHz; \citealt{Condon1998}), so our images of this GRB/SN suffered from strong artifacts and dynamic range issues. We intensively self-calibrated images to reach the noise thresholds listed in Table~\ref{MeasurementsTable}; note that the L-band data were much more severely affected by this source than the C-band data.

GRB\,$030329$/SN\,$2003$dh is surrounded by many sources, but none comparable in flux to the bright source in the GRB\,$060218$/SN\,$2006$aj images. Again, we self-calibrated our images to reach the noise thresholds listed in Table~\ref{MeasurementsTable}, and the L-band data were more severely affected by imaging artifacts than the C-band data.

A summary of the observations can be seen in Table~\ref{ObsTable}, and the results are listed in Table 3. Our observations were conducted 10 and 13 years after the initial explosions of GRB\,$060218$/SN\,$2006$aj and GRB\,$030329$/SN\,$2003$dh, respectively. Neither GRB/SNe were detected in either L or C band, so our observations provide 3$\sigma$ upper-limits on the flux densities (Table~3). These upper-limits are plotted on top of light curve models in Figures \ref{Fig_GRB03LightCurveModels} and \ref{Fig_GRB06LightCurveModels}.

\subsection{LBA Observations}
GRB\,980425/SN\,1998bw was observed with the LBA using an array comprising the Australia Telescope Compact Array (ATCA; phased array of five 22-meter dishes), Ceduna, Hobart and Parkes, on 16 November 2015 (Program ID V541A). The observing setup used 2$\times$16~MHz subbands in dual polarization, Nyquist sampled with 2~bits (256~Mbps data rate), and centred on a sky frequency of 1.65~GHz. A summary of the observations can be seen in Table~\ref{ObsTable}, and the results are listed in Table 3.

The observation had a duration of 12~hours, and GRB\,980425/SN\,1998bw was phase referenced to J1934$-$5053 which has 290~mJy of unresolved flux at a separation of $2^{\circ}$ from the target. Fringe finders 3C273, 1921$-$293 and 0208$-$512 were regularly observed to provide delay calibration; and a compact source, 1519$-$273, was used to bootstrap the intra-array flux calibration. The calibrator J1923$-$5329, located 3$^{\circ}$ from J1934-5053, was observed in a few scans in a phase referencing style similar to that used for GRB\,980425/SN\,1998bw, to confirm that phase transfer over a few degrees was successful.

All four Stokes parameters were correlated and, after calibration overheads, an on-source time of approximately 5 hours and 20 minutes on GRB\,980425/SN\,1998bw was achieved. Due to a partial disk failure at ATCA, about 35\% of the data from that station, randomly distributed throughout the experiment, were lost prior to correlation. The data were correlated on the LBA DiFX correlator \citep{Deller2011} and calibrated in NRAO's AIPS package in the standard way for LBA phase referencing using a pipeline implemented in the ParselTongue interface \citep{Kettenis2006}.

\begin{figure}[!t]
 \includegraphics[trim=0 0 0 0, clip, width=\columnwidth]{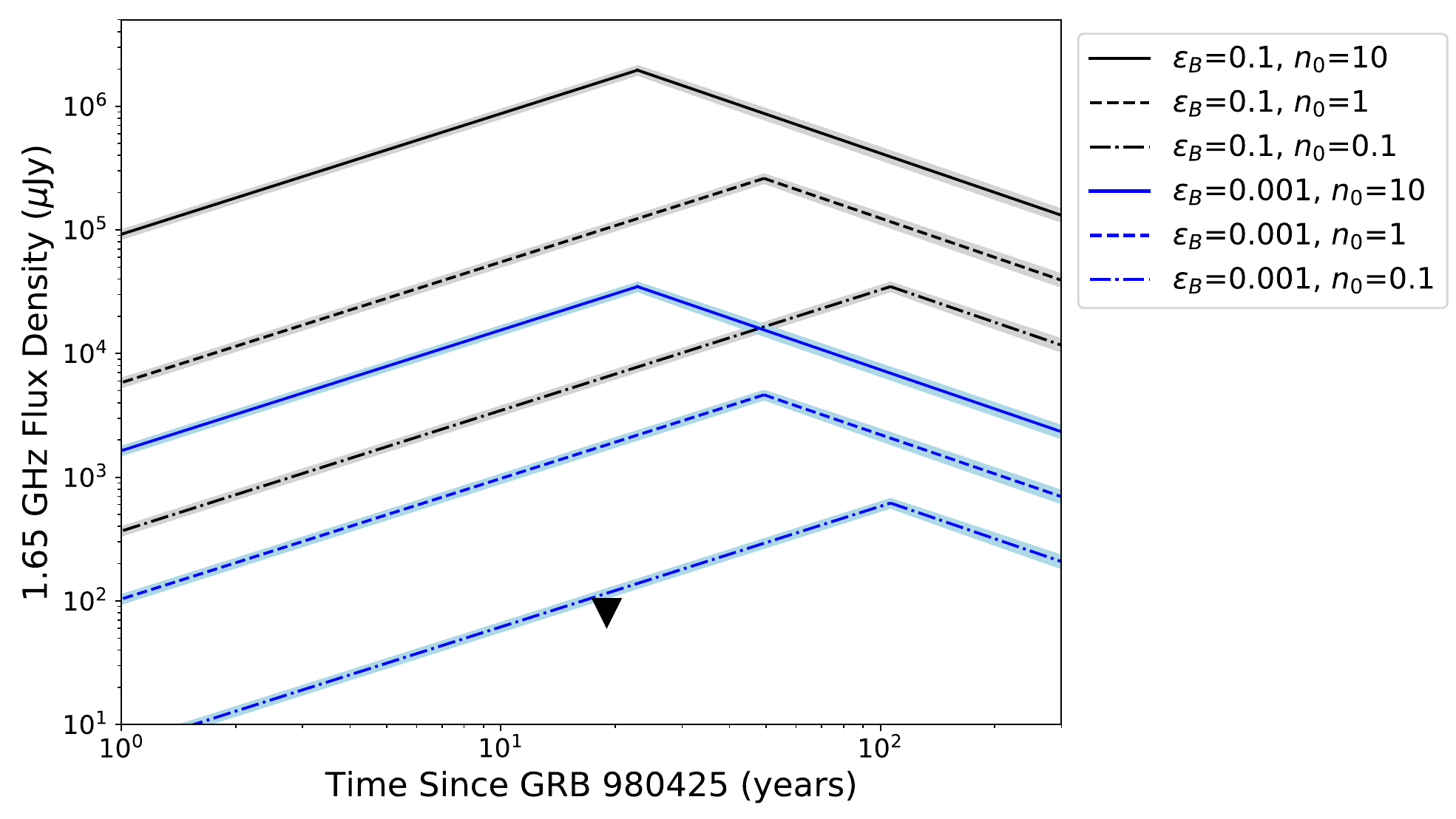}
 \caption{Model radio light curves of GRB\,980425/SN\,1998bw at $1.85$ GHz, with our observational limit over-plotted (black triangle). The light curve models, assumptions, and symbols are the same as for Figure 1, except we use the SN energy and velocity of SN\,1998bw from Table 1. Observations of the GRB\,980425 afterglow are not shown, as the GRB flux decreased rapidly and was not observed over the timescales featured in this plot.}
\label{Fig_GRB98LightCurves}
\end{figure}

The resultant naturally-weighted image noise was 90~${\mu}$Jy/beam. GRB\,980425/SN\,1998bw was not detected, giving a 3-$\sigma$ upper limit of $0.27$~mJy, 17.6 years after the initial explosion. Analysis of the quality of phase transfer from the check source, J1934$-$5053, indicate that phase calibration was good and thus the non-detection can be safely ascribed to weakness of the source rather than instrumental issues. This upper limit is the lowest limit published for GRB\,980425/SN\,1998bw \citep[cf.][]{Michalowski2012}, and shown on top of light curve models in Figure \ref{Fig_GRB98LightCurves}.

\begin{figure*}[!t]
 \includegraphics[trim=0 0 0 0, width=7in]{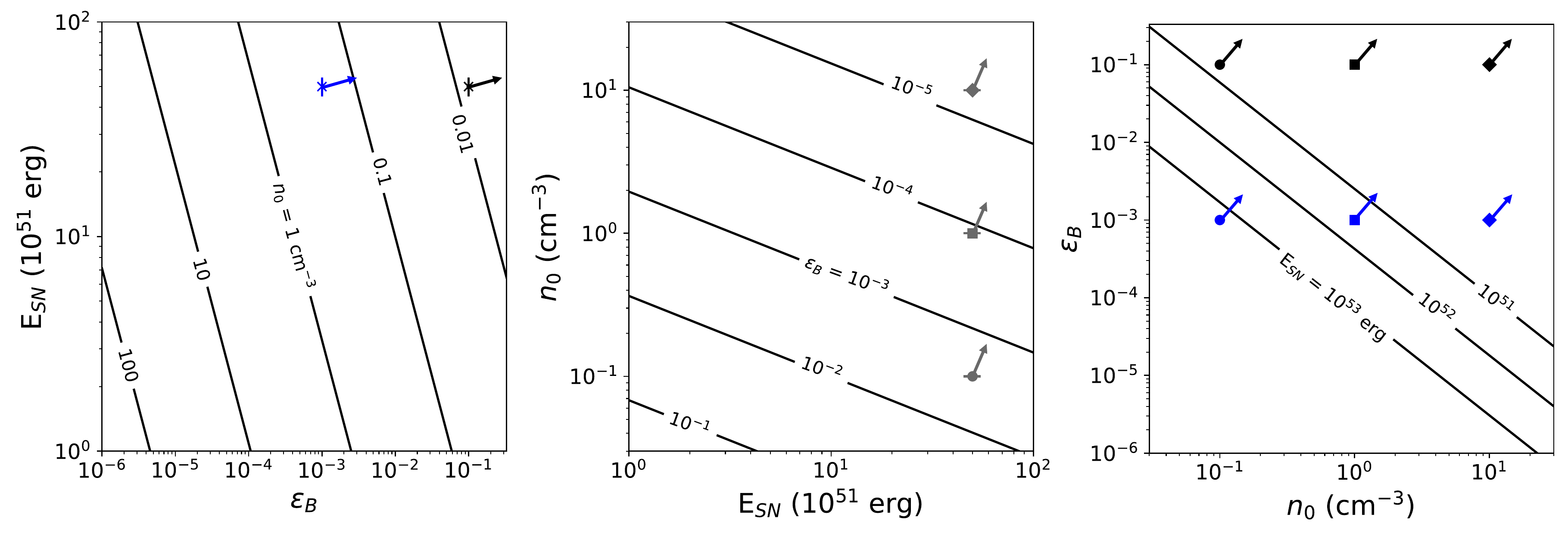}
 \caption{Given our radio upper limit on GRB\,$980425$/SN\,$1998$bw, 17.6 years after explosion, we can constrain the possible parameter space of $\epsilon_B$, $n_0$, and $E_{\rm SN}$ using Equation \ref{eps_vs_n}. The three panels illustrate how these three parameters depend on each other, assuming $p=2.5$, $\alpha=5$, $\epsilon_{e}=0.1$, and $v=24,000$~km~s$^{-1}$. The points on the plot refer to the fiducial values we assume in our previous figures. Color of the markers represents values of $\epsilon_{B}$, where black symbols correspond to $\epsilon_{B}=0.1$, blue symbols to $\epsilon_{B}=0.001$, and gray symbols assume nothing about $\epsilon_{B}$. The shape of markers corresponds to the values of $n_{0}$, where circles are $0.1$~cm$^{-3}$, squares are $1$~cm$^{-3}$, diamonds are $10$~cm$^{-3}$; and crosses are assuming nothing about the density. The arrows attached to each marker point to the regions of the parameter space allowed by our radio upper limits. For example, the square marker in the second panel assumes $n_{0}=1$~cm$^{-3}$ and $E_{\rm SN}=10^{52.5}$~erg (the energy for SN\,$1998$bw in Table 1); our radio upper limit therefore implies $\epsilon_{B}\lesssim 10^{-4}$ for those values of $n_{0}$ and $E_{\rm SN}$.}
\label{Fig_GRB98SetFluxPlots}
\end{figure*}

\section{Analysis} \label{s:5}

Here, we interpret our radio upper limits in the context of the radio rebrightening during the transition to the Sedov-Taylor phase. We note that non-detections of these GRB/SN sources also imply non-detections of GRB counter jets (the light emitted from a GRB jet that is expelled in a direction away from an observer, which means that the light will reach an observer at a later date than the jet expelled in the direction of the observer). We also conclude that the GRB radio afterglow is no longer detected at GHz frequencies for any of the three GRB/SN---even for GRB\,$030329$, where the afterglow was traced for almost a decade (Figure \ref{Fig_GRB03LightCurveModels}). Our observation of GRB\,$030329$/SN\,2003dh was made almost five years after the last of these observations, and our non-detection presented here demonstrates that any future radio detection of GRB\,$030329$/SN\,2003dh at GHz frequencies will likely originate from the SN ejecta rather than the GRB afterglow.

\subsection{Interpreting Radio Upper Limits} \label{s:5.1}

Figures \ref{Fig_GRB03LightCurveModels}, \ref{Fig_GRB06LightCurveModels} and \ref{Fig_GRB98LightCurves} show model light curves (described in \S \ref{s:2}) for each of the target GRB/SNe and our observational limits on flux density. In each of these figures we vary $\epsilon_B$ and n$_{0}$, while keeping the other model parameters fixed. For $\epsilon_B$ we adopt values of $0.1$ and $10^{-3}$---the former being equal to the assumed value for $\epsilon_e$ (i.e., equipartition) and the latter in the range of values that has been derived from GRB afterglow modeling \citep[e.g.,][]{Granot2014}. For GRB afterglows, it has been shown that $\epsilon_e$ is fairly narrowly distributed around 0.1 \citep[e.g.,][]{Beniamini2017}, and $\epsilon_e = 0.1$ is also commonly used for SNe Ib/c \citep{Chevalier_Fransson2006}. We note that for the slower shocks in ``normal'' SNRs (with velocities around a few thousand km s$^{-1}$), $\epsilon_e$ and $\epsilon_B$ can be one to two orders of magnitude lower than assumed here \citep[e.g.,][]{Sarbadhicary2017}. The ejecta of the GRB/SNe have high kinetic energies and should therefore maintain large velocities (around an order of magnitude faster than the velocities seen in typical SNRs), even as they transition to the Sedov-Taylor phase \citep{Kathirgamaraju2016}. Therefore, we take $\epsilon_e = 0.1$, but upper limits can be interpreted with other values of $\epsilon_e$ using Equation 6. For lower $\epsilon_e$ or $\epsilon_B$, we place less stringent constraints on the ambient density or SN energy with our radio upper limits (Equation~\ref{eps_vs_n}, Figure~4).

For $n_0$ we adopt values of $0.1$, $1$ and $10$ cm$^{-3}$, spanning a range of most common values found in GRB afterglow modeling \citep{Granot2014}. We can use our radio upper limits to constrain the density of the ambient medium surrounding a GRB/SN. Figures \ref{Fig_GRB03LightCurveModels}, \ref{Fig_GRB06LightCurveModels} and \ref{Fig_GRB98LightCurves} demonstrate how variations in $n_0$ affect the radio luminosity. We see that the density of the surrounding medium plays a large role in not only the Sedov-Taylor time but also the light curve peak flux.

The predicted radio light curves also depend on the energy and ejecta velocity of the SN. With higher velocities, we expect that the radio luminosity would peak at earlier times and higher luminosities (Equations 1 and 2). For larger ejecta masses, the radio luminosity would peak at later times and higher luminosities. While ejecta velocities are well constrained by observations, measurements of ejecta mass and E$_{SN}$ are model-dependent and have substantial uncertainty, with possible values for SN\,$2003$dh ranging by an order of magnitude (Table \ref{ObsTable}). We take the SN energies listed in Table \ref{SampleTable} as fiducial parameters, but can quantify how our upper limits depend on E$_{SN}$ (Figure \ref{Fig_GRB98SetFluxPlots}). Note that we assume $p = 2.5$ and $\alpha = 5$, which are both typical for SN modeling \citep{Chevalier_Fransson2006}.

Clearly, the light curves depend on multiple uncertain parameters, e.g., $\epsilon_e$, $\epsilon_B$, $n_0$, and E$_{SN}$. If we assume values for three of these parameters, we can then place clear constraints on the fourth. In Figure \ref{Fig_GRB98SetFluxPlots}, we use our $1.65$ GHz upper limit for GRB $980425$/SN $1998$bw to constrain the three more uncertain input parameters ($\epsilon_B$, $n_0$, and E$_{SN}$) and illustrate the degeneracies between them, while fixing the other parameters ($p = 2.5$, $\alpha = 5$, and $\epsilon_e = 0.1$). The three different panels demonstrate how a third parameter depends on the other two. For example, if a viewer of the the left panel of Figure 4 selected a value for $E_{\rm SN}$ and $\epsilon_B$, they could read off the value of the corresponding density contour to place an upper limit on $n_0$. The blue point in this panel marks fiducial values of $\epsilon_{B}=0.001$ and $E_{\rm SN} = 10^{52.5}$ erg, and lands between the $n_0 = 1$ cm$^{-3}$ and $n_0 = 0.1$ cm$^{-3}$ contours, implying that $n_0 \lesssim 0.3$ cm$^{-3}$ under these assumptions.

Although we do not include similar plots for GRB\,$030329$/SN\,$2003$dh and GRB\,$060218$/SN\,$2006$aj, the same ideas can be followed and we can calculate the range of expected densities for these sources. For these calculations, we turn to the C-band ($6.0$ GHz) observations, as they are more constraining than the L-band. We keep the same assumed parameter range for $\epsilon_{B}$ ($0.1$ and $0.001$) and $E_{\rm SN}$ is allowed a range that spans the uncertainty on the measurements from Table \ref{SampleTable}. For both GRB\,$030329$/SN\,$2003$dh and GRB\,$060218$/SN\,$2006$aj we find $n_0 \lesssim 0.3-9$ cm$^{-3}$.

\subsection{Implications for GRB/SN Environments} \label{s:5.2}

Our radio upper limits probe the density of the circumstellar material (CSM) or interstellar medium (ISM) at the location of the SN forward shock at the time of observation. By estimating the radii of the ejecta and using our upper limits to constrain $n_0$, we can comment on the environments of GRB/SNe.

We estimate the radius of the SN shock as $R_s = v t_{\rm obs}$, where $v$ is the SN ejecta velocity (Table 1) and $t_{\rm obs}$ is the time elapsed between explosion and observations (10.0--17.6 yr; Table 3). We find radii of $R_s\approx0.2$~pc for GRB\,$060218$/SN\,$2006$aj and $R_s\approx0.4$~pc for GRB\,$980425$/SN\,$1998$bw and GRB\,$030329$/SN\,$2003$dh. These radii are in fact lower limits, with the true blast-wave velocity a factor $\sim$3 larger, as the SN has not yet reached the Sedov-Taylor stage, implying that the velocity of the fastest SN ejecta are faster than $\beta_{\rm SN}$.

In order to place constraints on the density of the sources that we have observed, we turn to what environments are found around similar sources. Several authors have shown that there is a wide range of densities surrounding long GRBs, spanning many orders of magnitude \citep{Cenko2011,Granot2014}. \cite{Chandra2012} suggest that GRB radio samples are biased to a narrow range of CSM densities (1--10 cm$^{-3}$), as the radio emission will be weak at low densities and self-absorbed at high densities. In a fraction of cases, the radio light curves of long GRBs can be well-fit by expansion into a uniform medium, and this is what we assume here (Section \ref{s:2}). However, a uniform-density CSM is almost certainly over-simplistic, as the environments of GRB/SNe should be strongly affected by the evolution and mass loss from the progenitor star \citep[e.g.,][]{Ramirez2005,Starling2008}. 

A simple model for the progenitor's evolution in the years leading up to explosion might be a fast wind ($\sim$1000 km s$^{-1}$) sustained for $\sim 10^5$ years, as expected for a Wolf-Rayet star \citep{Crowther2007}. Such a progenitor should blow a bubble filled with a $\rho \propto r^{-2}$ wind, implying low densities at radii $\sim$0.1--10 pc \citep{Weaver1977}. More realistic stellar progenitors yield substantially more complex circumstellar environments. Take for example the 29 M$_{\odot}$ star whose late stages of evolution are modeled in Figure 3 of \cite{Ramirez2005}. The star is surrounded by a Wolf-Rayet wind-blown bubble, but the bubble is both smaller and denser than one might naively expect because it is expanding into a dense wind from the red supergiant phase that preceded the Wolf-Rayet phase. In this particular simulated CSM, the Wolf-Rayet bubble has a diameter of $\sim$0.3 pc and density $\sim$10 cm$^{-3}$, and is surrounded by a dense shell of material of density $\sim 10^2-10^4$ cm$^{-3}$. These examples highlight how difficult it is to predict the CSM around a GRB/SN. Even a question as simple as whether the ejecta are expanding into a medium that is enhanced in density over the ISM or evacuated of ISM is difficult to answer and depends on the detailed mass loss of the progenitor star. 

Adding to the complexity of the CSM, the SN blast waves studied here are expanding into a medium that has already been shaped by the lower-mass, higher-velocity GRB ejecta. For example, the afterglow of GRB\,$030329$ is best modeled by interaction with a uniform density medium, with $n_0$ values ranging about an order of magnitude around 1~cm$^{-3}$ out to a radius $\gtrsim $1 pc \citep[e.g.,][]{vanderHorst2008,Mesler2013}. The SN ejecta in GRB $030329$/SN $2003$dh has a radius $\lesssim$0.8 pc at the time of our observations, 13 years after explosion. Therefore, our observations probe the SN blast wave while it is interacting with the GRB-evacuated cavity. Figure \ref{Fig_GRB03LightCurveModels} shows that for equipartition with $\epsilon_e=\epsilon_B=0.1$ the density should be below 1~cm$^{-3}$, consistent with only some of the density values found in broadband modeling of the GRB emission. For $\epsilon_e=0.1$ and $\epsilon_B=10^{-3}$ the density is less well constrained, $n_0<10$~cm$^{-3}$, which is consistent with all the density values derived from modeling GRB\,$030329$.


The radio afterglow of GRB\,$060218$ can be fit with a stellar wind density profile or a uniform density medium of $n_0 \approx 10^2$ cm$^{-3}$ \citep{Soderberg2006, Toma2007, Irwin2016}. Our upper limit shown in Figure \ref{Fig_GRB06LightCurveModels} provides constraints on the density similar to those for GRB\,$030329$: $n_0<1$~cm$^{-3}$ for $\epsilon_e=\epsilon_B=0.1$, and $n_0<10$~cm$^{-3}$ for $\epsilon_e=0.1$ and $\epsilon_B=10^{-3}$. Those limits on $n_0$ are still below the value derived from modeling the GRB\,$060218$ afterglow with a uniform density medium. This may indicate that the density of the CSM drops with radius, implying a stellar wind density profile, which means that our upper limit may be breaking the degeneracy between possible density profiles for modeling this GRB afterglow. We emphasize that this depends on the $\epsilon_e$ and $\epsilon_B$ values, which can both be lower than the ones we show in Figure \ref{Fig_GRB06LightCurveModels}.

Finally, the radio emission from SN\,1998bw is best modeled by interaction with a stellar wind rather than a uniform CSM \citep{Li1999, Weiler2001}. Predictions for the SNR emission associated with a GRB/SN expanding into a wind CSM are outside the scope of this paper. However, Figure \ref{Fig_GRB98LightCurves} shows strong constraints on the density: even for $\epsilon_e=0.1$ and $\epsilon_B=10^{-3}$, the upper limit on $n_0$ is $0.1$~cm$^{-3}$ at $R_s = 0.4$~pc; see the right panel of Figure \ref{Fig_GRB98SetFluxPlots} for the correlation between $\epsilon_B$ and $n_0$. This would be consistent with the density expected for a stellar wind with mass loss rate of $\dot{M} = 6 \times 10^{-7}$ M$_{\odot}$ yr$^{-1}$ for an expansion velocity $v_w = 1000$ km s$^{-1}$ \citep{Li1999}. This mass loss rate is quite low for typical Wolf-Rayet stars and mass loss rates derived from GRB modeling, although not unprecedented for the latter \citep{vanderHorst2014}.

\section{Conclusions} \label{s:6}
In this paper, we presented observations of three long GRBs with associated SNe in an effort to detect rebrightening radio emission from the SN ejecta entering the Sedov-Taylor phase. We observed GRB\,$030329$/SN\,$2003$dh and GRB\,$060218$/SN\,$2006$aj with the VLA, and GRB\,$980425$/SN\,$1998$bw with the LBA. Our observations resulted in non-detections, with $L_{\nu} \lesssim [0.4-10^2] \times 10^{27}$~erg~s$^{-1}$~Hz$^{-1}$. By choosing fiducial values for parameters describing the SN energetics and shock microphysics, we place upper limits on the density surrounding the GRB/SNe at radii $\sim$0.2--0.8 pc from the explosion site. 

We find that the density limits for GRB\,$030329$/SN\,$2003$dh are similar to the density values derived from afterglow modeling, while the limits for GRB\,$060218$/SN\,$2006$aj and GRB\,$980425$/SN\,$1998$bw are quite low. For GRB\,$060218$/SN\,$2006$aj, the limits on $n_0$ may break the degeneracy between possible density profiles for modeling the GRB afterglow, i.e. they prefer a stellar wind profile over a homogeneous CSM, unless $\epsilon_e$ and $\epsilon_B$ are significantly below $0.1$ and $10^{-3}$, respectively. In the case of GRB\,$980425$/SN\,$1998$bw, the limits on the density imply a low but not unprecedented mass loss rate of the progenitor's stellar wind.

While our observations resulted in non-detections, our upper limits are ruling out significant fractions of parameter space for some of the physical parameters of GRB/SNe. A future detection of the SNR emission from decades-old GRB/SNe will enable a better understanding of the environments of long GRBs and illuminate the transition from SN to SNR. 



\acknowledgments
This work was supported by the University of Wisconsin--Madison's Advanced Opportunities Fellowship (AOF), National Science Foundation Grants  AST-$1412549$, AST-$1412980$ and AST-$1413099$, and a Scialog grant from the Research Corporation for Science Advancement. C.~P.\ thanks the LSSTC Data Science Fellowship Program, her time as a Fellow has benefited this work. C.~P.\ was supported by NASA under Award No. RFP$17$\_$6.0$ \#NNX15AJ12H issued through Wisconsin Space Grant Consortium. RBD and DG acknowledge support from the National Science Foundation under Grants $1816694$ and $1816136$.

The National Radio Astronomy Observatory is a facility of the National Science Foundation operated under cooperative agreement by Associated Universities, Inc. The Long Baseline Array is part of the Australia Telescope National Facility which is funded by the Australian Government for operation as a National Facility managed by CSIRO. This work was supported by resources provided by the Pawsey Supercomputing Centre with funding from the Australian Government and the Government of Western Australia.

\bibliographystyle{mnras}
\bibliography{EarlyGRBSN.bib}

\end{document}